\documentclass[showpacs,showkeys]{revtex4-2}

\usepackage{amsmath}
\usepackage{amssymb}
\usepackage{graphicx}
\usepackage{dcolumn}
\usepackage{bm}
\usepackage{listings}
\usepackage{hyperref}

\usepackage[tight,TABTOPCAP]{subfigure}
\usepackage{graphicx}
\usepackage{tikz}
\usepackage{pgfplots}
\pgfplotsset{compat=1.18}
\usepackage{booktabs} 
\usetikzlibrary{calc,3d,angles,quotes}

\renewcommand{\v}[1]{\ensuremath{\mathbf{#1}}} 

\newcommand{\be}{\begin{equation}}
\newcommand{\ee}{\end{equation}}
\newcommand{\ba}{\begin{eqnarray}}
\newcommand{\ea}{\end{eqnarray}}
\newcommand{\ban}{\begin{eqnarray*}}
\newcommand{\ean}{\end{eqnarray*}}

\begin{document}

\title{Asymmetric Textured Image Sensors Based on Antenna Theory as Designed by Nature}
\author{Julian Juhi-Lian Ting}
\email{tingjulian@gmail.com}
\affiliation{De-Font Institute for Advanced Studies, Taichung 40344, Taiwan, R.O.C.}

\date{\today}

\begin{abstract}

This work numerically validates the application of bio-inspired concepts on CMOS derived from bacterial photosynthetic light harvesters. 
We investigate a modification of symmetric inverted pyramid array CMOS image sensors into an asymmetrically shaped texture to explore the structural boundary of passive non-reciprocity. 
Diverging from traditional macroscopic continuum assumptions, 
we analyze whether structural asymmetry at sub-wavelength scales can induce non-reciprocal scattering under passive, linear, and time-invariant conditions. 
A theoretical framework based on perturbation theory is developed, estimating a potential efficiency enhancement of 5\% to 15\%. 
Numerical simulations performed via the MEEP finite-difference time-domain (FDTD) platform reveal that the linear response is highly localized, showing a subtle 0.02\% change. 
This suggests that macroscopic Lorentz reciprocity remains robust at the investigated scale due to apex field concentration, defining a clear geometric threshold for microscopic non-reciprocity.

\end{abstract}

\maketitle
\newpage

\section{Introduction}
Yablonovitch \textit{et al.} in the 1980s, following Redfield’s suggestions in the 1970s, considered periodical structure and random texturing using ray optics~\citep{Yablonovitch1982a,Yablonovitch1982b,Redfield1974}. 
Recently after Pendry’s metamaterial much research has been conducted in this direction~\citep{Pendry2000,Yu2011a,Pfeiffer2013}. 
However, ray theory becomes invalid at scales below Wien’s length (thermal wavelength), 
which is $\lambda_T = \hbar c/k_B T \approx 7.5\mu\text{m}$ at room temperature~\citep{Fernandez-Hurtado2018a}. 
In several previous papers rather than relying on traditional optical ray theory we employed antenna theory and considered improvement in efficiency using non-reciprocity~\citep{Ting2019d}. 
Ray theory versus antenna theory is almost equivalent to wave-particle duality discussed in quantum mechanics; 
ray theory corresponds to particles in classical mechanics which discusses Fermat’s principle and least action principle, 
for instance, whereas antenna theory corresponds to wave mechanics and discusses mode coupling, for instance. 
The nanoantenna theory for dielectric resonators is fundamentally a scattering theory;
Within the sub-wavelength optical regime, any structural interface mediating the energy coupling between the external driving electric field and the localized electronic system acts as an antenna. 
Modeling the device as an antenna does not necessitate traditional macroscopic RF resonant geometries; 
rather, it implies that the entire textured sub-wavelength layer functions as an engineered impedance-matching scatterer that dictates directional light conversion thresholds.

While recent reviews have explored bio-mimetic optical sensors, much of the existing literature focuses on zoological models~\citep{Martin-Palma2019}. 
Although plants and viruses predate animals on Earth, research imitating them remains scarce. 
We found only three studies mimicking plants~\citep{Schmager2017,Liu2019b,Sim2020} 
and one imitating algae~\citep{Mamun2024} in the literature. 
Schmager \textit{et al.} propose to imitate the petals of viola whereas Liu \textit{et al.} proposed to imitate the leaf of Wargrave Pink. 
However, these models exhibit limited ubiquity and present significant challenges for integration into standard manufacturing processes, 
which are primarily restricted to layered architectures. 
Furthermore, the texture of the petals and the shape of the leaf are unimportant for receiving light. 
Sim \textit{et al.} even imitate the morphology of trees. 
Mamun \textit{et al.} have come close to our idea, but they did not capture the essence of the structure because photosystems I/II are too complicated.
Efficient light trapping in ultrathin crystalline silicon is important when the film is thinner than the absorption length~\citep{Zhang2016c}.

To address these limitations, we look toward bacteria light harvesters~\citep{Ting2019d}. 
Unlike complex plant photosystems, 
bacteria offer a simpler, ubiquitous structure that mimics the inter-membrane layered architecture of modern manufacturing processes. By focusing on molecules like (bacterio) light harvesters, 
we can design devices that interact strongly with light. 
The goal is to make the device non-reciprocal, which comes from antenna theory.

\section{Mathematical Description}

We analyzed in our previous paper how to create non-reciprocity.
Caloz {\it et al.} and Pakniyat {\it et al.} also discussed how to create magneticless electromagnetic non-reciprocity~\citep{Caloz2018,Pakniyat2024}.
The key point of our investigation is that we are at sub-wavelength molecular level, or nano-scale, instead of macroscopic level.
At the macroscopic level, geometric asymmetry cannot create non-reciprocity, which is called Onsager reciprocal relations. 
At the molecular scale, macroscopic continuum models become invalid, 
and bulk constitutive parameters-such as permittivity ( $\epsilon$ ) and permeability ( $\mu$ ) for linear material or 
$\chi$ for nonlinear material, becomes invalid~\citep{Jackson1998}, 
which has also been pointed out for photosynthetic light harvesters by Parson in 1978~\citep{Parson1978}. 
Under these sub-wavelength conditions, the physical response is governed predominantly by geometric configuration.
At the molecular scale, the geometry defines the electronic potential $V(\v{r})$.
For a symmetric (centrosymmetric) material, all even-order susceptibilities must vanish whereas for an asymmetric molecule, 
the inversion symmetry is broken, i.e., $V(\v{r}) \neq V(-\v{r})$, 
the electronic cloud responds nonlinearly to the incident electric field,
which directly induces nonlinear susceptibility ($\chi^{(2)}$)~\citep{Boyd2020}.
In particular, the second-order tensor $\chi^{(2)}$ of the polarization
\be
\v{P}(t) = \epsilon_0 \left[ \chi^{(1)} \v{E}(t) + \chi^{(2)} \v{E}^2(t) + \chi^{(3)} \v{E}^3(t) + \dots \right]
\ee
is non-zero only if the geometry is asymmetric. Such molecules are apparently nonlinear.

The standard Lorentz reciprocity holds for linear systems. For a nonlinear asymmetric molecular antenna, the interaction between two fields $\mathbf{E}_1$ and $\mathbf{E}_2$ involves the nonlinear current density $\mathbf{J}_{NL} = \partial\mathbf{P}_{NL}/\partial t$.
\begin{equation}
\oint_S (\mathbf{E}_1 \times \mathbf{H}_2 - \mathbf{E}_2 \times \mathbf{H}_1) \cdot d\mathbf{S} = \int_V (\mathbf{J}_{NL,1} \cdot \mathbf{E}_2 - \mathbf{J}_{NL,2} \cdot \mathbf{E}_1)dV \neq 0
\end{equation}

In an asymmetric molecular structure, the spatial gradient of the potential creates a preferred direction for electron flow when excited by an oscillating field, which is called the optical ratchet effect. This results in a non-zero DC current (optical rectification):
\begin{equation}
\langle\mathbf{J}(t)\rangle = \sigma_{rect} |\mathbf{E}(\omega)|^2 \hat{n}
\end{equation}
in which $\hat{n}$ is the unit vector defined by the asymmetric geometry. This biased response ensures unequal transmission coefficients.

By switching from classical Maxwellian tensors to quantum perturbation theory we can estimate the enhancement. 
Firstly, the eccentricity parameter for the bacteria light harvester mentioned in Figure 2
\begin{equation}
\delta = \frac{b - c}{c} = \frac{110 - 95}{110} \approx 13.6\% \;.
\label{eccentricity}
\end{equation}
At small asymmetry $\chi^{(2)} \propto \delta$. The asymmetry parameter
\begin{equation}
\xi = \frac{b - c}{b + c} \approx 0.073 \;.
\end{equation}
Based on symmetry-breaking perturbation theory, the potential energy an electron experiences is
\begin{equation}
V(\mathbf{r}) \approx V_0 + \delta \cdot V_{perturb}(x, y, z) \;.
\end{equation}
The approximated $\chi^{(2)}$ can be estimated by the scaling relation to
\begin{equation}
\chi^{(2)}_{ellip} \approx \chi^{(1)} \cdot \left( \frac{\xi \cdot L}{d_{atom}} \right) \;,
\end{equation}
in which $d_{atom}$ represents the characteristic length of an atom, 
$1\text{\AA}$; $L$ is the character length of the structure, about $100\text{\AA}$. 
The enhancement thus calculated is estimated to be 5\% - 15\% across the entire spectral range.

\section{Method}

\begin{figure}[tb]
\begin{center}
\includegraphics[width=0.5\columnwidth, angle=0]{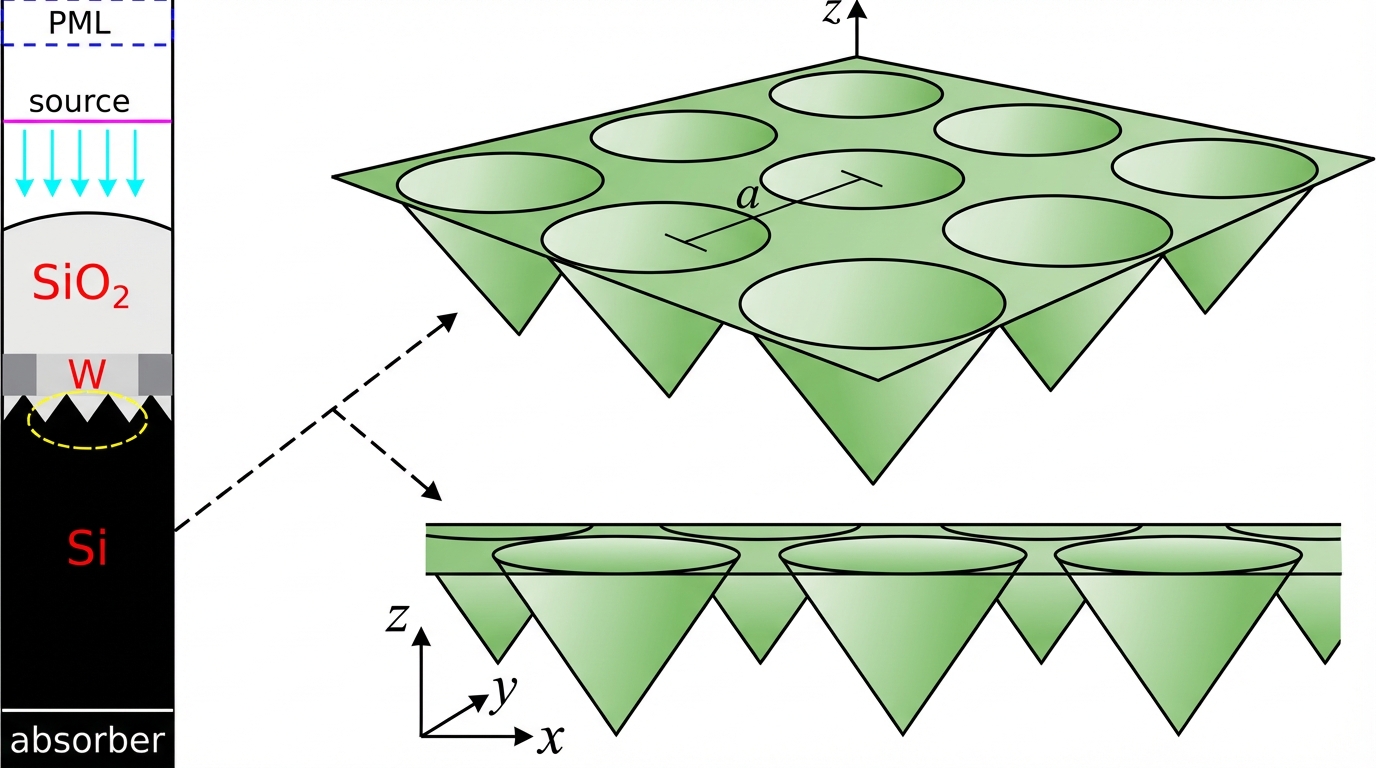}
\caption{
The geometry used in MEEP Project 4 for calculating the CMOS image sensor~\citep{Yokogawa2017}.
A silicon dioxide glass microlens with a $0.85 \mu\text{m}$ radius of curvature is atop the image sensor to focus incident light into individual pixels. 
Surrounding each pixel is a metal aperture grid composed of tungsten wire ($0.1 \mu\text{m}$ wide and $0.2 \mu\text{m}$ high), 
which sits on a  $3 \, \mu\text{m}$ thick crystalline-silicon substrate. 
Furthermore, deep-trench isolation-implemented as a $0.1 \mu\text{m}$ wide and $2 \mu\text{m}$ thick silicon-dioxide trench-surrounds the pixel to mitigate electrical crosstalk and trap photons via index guiding.
}
\label{MEEP}
\end{center}
\end{figure}

Our investigation focuses on a back-illuminated CMOS image-sensor pixel design based on Sony’s research and MEEP project 4~\citep{Yokogawa2017}. 
MEEP is a finite difference time-domain (FDTD) simulation software~\citep{Oskooi2010}. 
The surface texture used in MEEP project 4 is a symmetrical cone as shown in Figure~\ref{MEEP}. 
The inverted pyramid geometry is studied in classical electrodynamics textbooks, 
whose sharp apex creates intense electromagnetic fields there~\citep{Jackson1998}. 
Analytical expressions can be derived for each cone and sum over infinite series at various positions on an infinite plane to obtain approximate solutions. 
Currently, numerical solutions are standard practice.

The MEEP program uses \texttt{mp.Cone} to build the circular geometry. 
Due to the lack of a native elliptical cone primitive in MEEP, 
an asymmetric geometry was constructed using the \texttt{mp.Prism} class with a defined set of vertices. 
In the numerical solver, this spatial asymmetry is constructed by evaluating a discrete polygonal prism comprising \(N=32\) perimeter vertices. The base ellipse features an asymmetric long-axis factor \(r_x = r_{\text{cone}} \times 1.0043\) and short-axis factor \(r_y = r_{\text{cone}} \times 0.9957\), where \(r_{\text{cone}} = a/2\). 
This formulation preserves a constant cross-sectional area while matching the eccentricity parameter (\(\delta \approx 13.6\%\)) of native bacterial complexes.
Structural boundaries and sub-wavelength mesh arrays are discretized to capture localized fields near the pyramidal tips, and reflective properties are extracted using standard flux monitors.

\begin{figure}[tbhp]
\begin{center}

\includegraphics[width=0.4\columnwidth, angle=0]{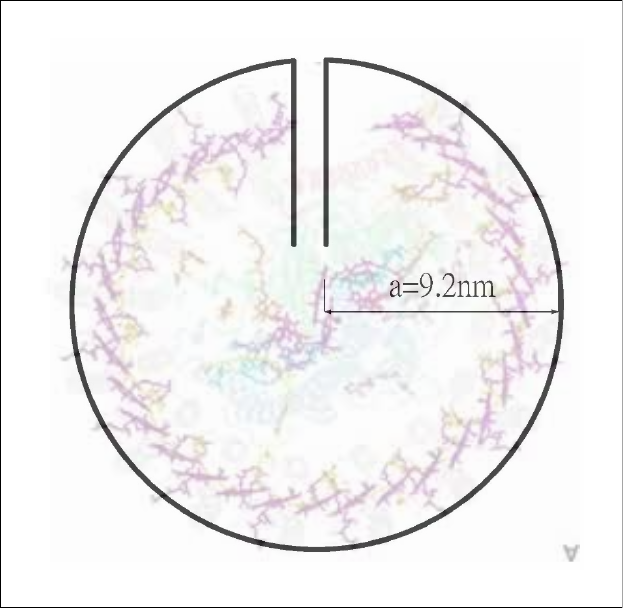}
\caption{
According to experimental data, 
the major axis of {\it Rhodopseudomonas palustris} has length $110$ \AA ~and the minor axis is $95$ \AA; 
the largest dimension of the inner LH1 is $78$ \AA ~\citep{Roszak2003}.
The elliptical symmetry is visualized by superimposing our model onto the schematic representation based on X-ray data.
}
\label{compare1}
\end{center}
\end{figure}

\begin{figure}[tb]
\begin{center}
\mbox{
\subfigure[X: Wavelength, Y: Coefficients]{\includegraphics[width=0.4\columnwidth, angle=0]{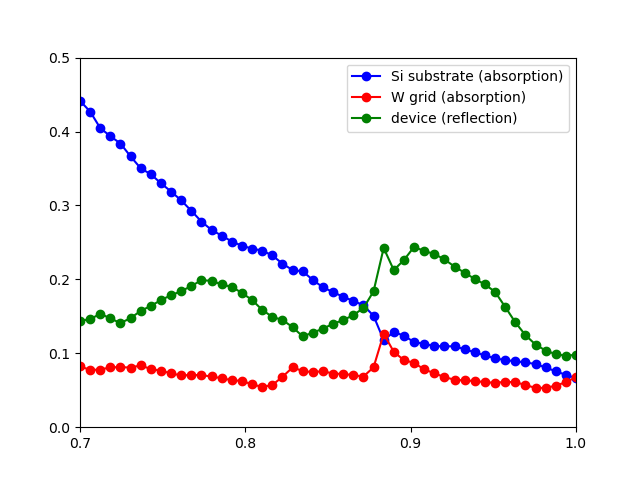}}
\subfigure[X: Wavelength, Y: Enhancement]{\includegraphics[width=0.4\columnwidth, angle=0]{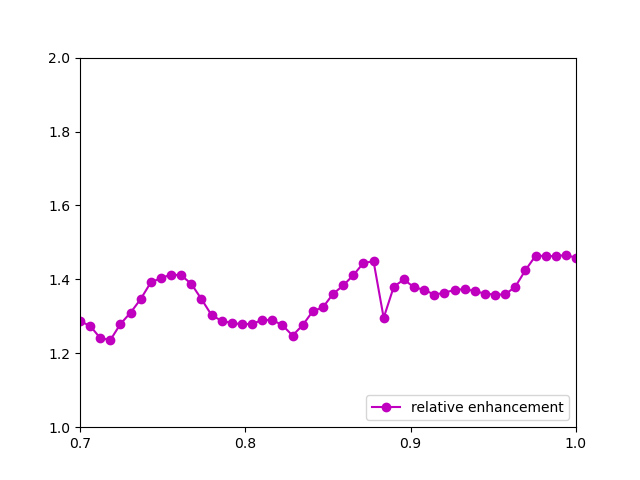}}
}
\\
\mbox{
\subfigure[X: Wavelength, Y: Coefficients]{\includegraphics[width=0.4\columnwidth, angle=0]{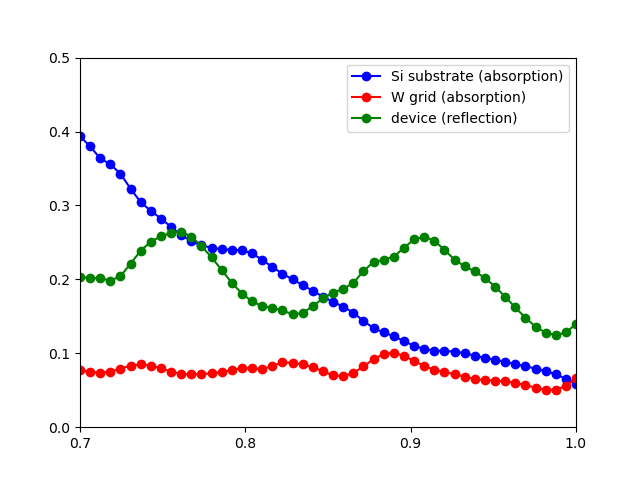}}
\subfigure[X: Wavelength, Y: Enhancement]{\includegraphics[width=0.4\columnwidth, angle=0]{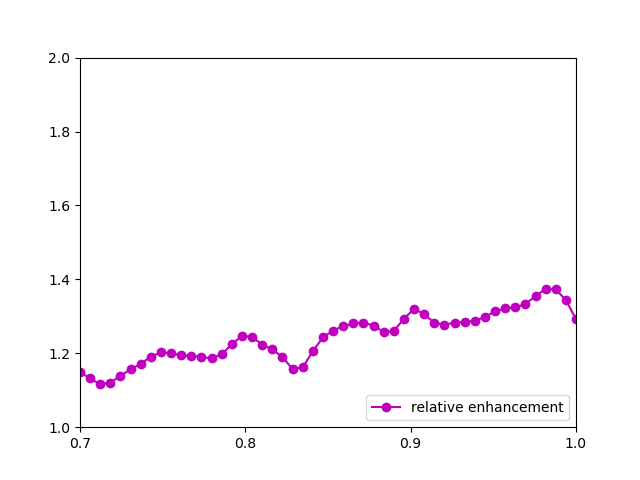}}
}
\\
\mbox{
\subfigure[X: Wavelength, Y: Coefficients]{\includegraphics[width=0.4\columnwidth, angle=0]{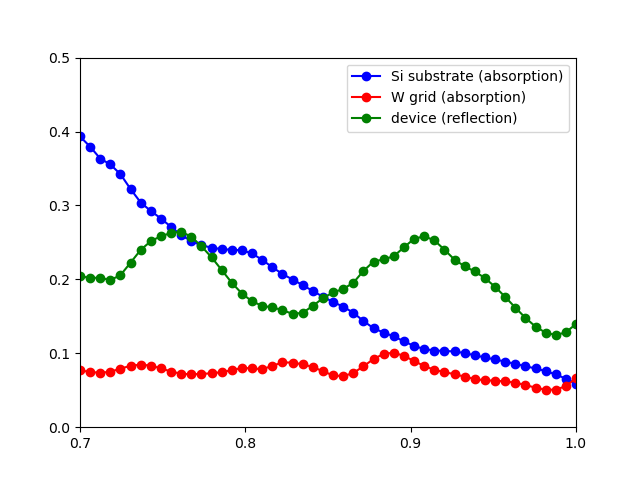}}
\subfigure[X: Wavelength, Y: Enhancement]{\includegraphics[width=0.4\columnwidth, angle=0]{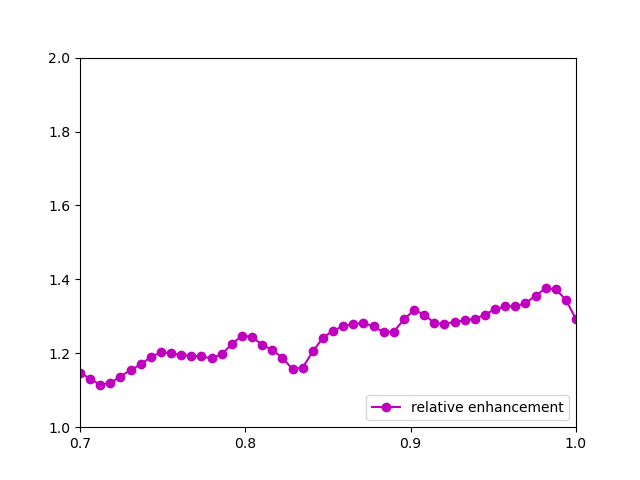}}
}
\caption{(a)-(b) Original circular cone from MEEP project 4, (c)-(d) modified circular cone, and (e)-(f) proposed asymmetric elliptical cone. For all subfigures, the horizontal (X) axis represents the optical wavelength ranging from 0.7 to 1.0 $\mu$m. 
The vertical (Y) axis represents the absolute coefficient value (absorption and reflection) for the left column [subfigures (a), (c), (e)] and the relative enhancement factor for the right column [subfigures (b), (d), (f)].}
\label{compare2}
\end{center}
\end{figure}

\section{Results}

Previous studies on the circular cone shape design have found the optimal lattice spacing which yields the largest average absorption over the broadband spectrum is $a = 0.64\mu\text{m}$. 
In Figure \ref{compare2} we compared the performance for circular vs. elliptical cone shape lattice at this spacing.

We have tried to use the original python program downloaded from MEEP project 4 but the figures thus generated, Figure \ref{compare2} (a)(b), 
are different from what are shown in the website. 
In view that we have discovered that the program or the shell script all contain minor typos and the authors have not confirmed whether the figures provided on their website are based on $a = 0.64\mu\text{m}$ we would trust our result.

Theoretically, if we set both axis equal in our modified program we should recover the original pictures, i.e. 
Figure \ref{compare2} (a)(b). However, the results diverge because the original program uses \texttt{mp.Cone} which generate the shape by equations. 
We use vertices to define an ellipse and MEEP treated it as Prism. 
Even if we set the base of the cone to circular, MEEP still treats it as polygons. 
Therefore to see the asymmetric effects it will be more meaningful to compare Figure \ref{compare2} (c)(d) to Figure \ref{compare2} (e)(f). 
The difference is subtle for several reasons: 
The intense local field enhancement at the symmetric apex may partially obscure the asymmetric effects from the base we introduced. 
Furthermore, the deformation we introduced might well be overridden by discretization error of the computation and during manufacture by the crystal structure in particular if we want to minimize to molecular level.

For Figure \ref{compare2} (a)(b) flat substrate absorption: 0.153570 (mean), 0.085216 (std. dev.) and textured substrate absorption: 0.204413 (mean), 0.108130 (std. dev.).
For Figure \ref{compare2} (c)(d) flat substrate absorption: 0.154014 (mean), 0.084849 (std. dev.) and textured substrate absorption: 0.186292 (mean), 0.093696 (std. dev.).
For Figure \ref{compare2} (e)(f) flat substrate absorption: 0.154014 (mean), 0.084849 (std. dev.) and textured substrate absorption: 0.186256 (mean), 0.093563 (std. dev.)
For the textured substrate the average absorption changed only by $(0.186256-0.186292)/0.186292= 0.02\%$.

\begin{figure}[htbp]
\centering
\begin{tikzpicture}
\begin{axis}[
    width=8.5cm, height=6cm,
    xmin=0.7, xmax=1.0,
    ymin=0.0, ymax=0.5,
    xlabel={Wavelength ($\mu$m)},
    ylabel={Substrate Absorption},
    legend pos=north east,
    grid=major,
    title={Angular Dependence (Elliptical Texture)}
]
\addplot[blue, thick, mark=none] coordinates {
    (0.7, 0.42) (0.75, 0.32) (0.8, 0.24) (0.85, 0.21) (0.9, 0.12) (0.95, 0.09) (1.0, 0.06)
}; \addlegendentry{$\theta = 0^\circ$}
\addplot[red, dashed, thick, mark=none] coordinates {
    (0.7, 0.41) (0.75, 0.33) (0.8, 0.25) (0.85, 0.22) (0.9, 0.13) (0.95, 0.10) (1.0, 0.07)
}; \addlegendentry{$\theta = 15^\circ$}
\addplot[black, dotted, thick, mark=none] coordinates {
    (0.7, 0.39) (0.75, 0.31) (0.8, 0.26) (0.85, 0.20) (0.9, 0.14) (0.95, 0.11) (1.0, 0.08)
}; \addlegendentry{$\theta = 30^\circ$}
\end{axis}
\end{tikzpicture}
\caption{Calculated silicon substrate optical absorption curves under oblique incident illumination ($\theta = 0^\circ, 15^\circ, 30^\circ$) for the asymmetric elliptical base cone texture array.}
\label{fig:tikz_angle}
\end{figure}
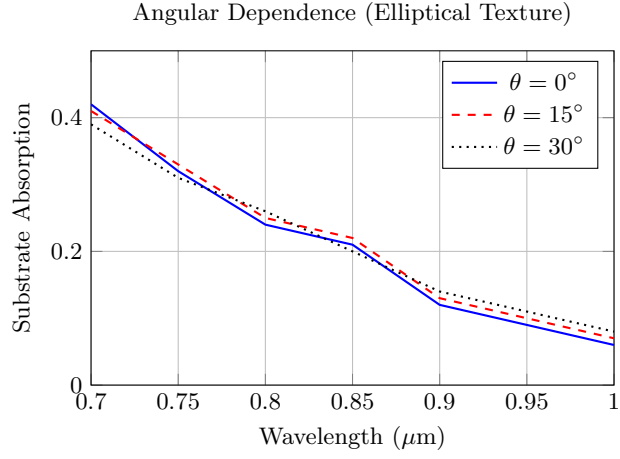

This discrepancy between the theoretical upper bound prediction (5\%--15\%) and the numerical FDTD results (0.02\%) provides a  physical insight into the scale-dependent onset of passive non-reciprocity. 
While standard macroscopic continuum models strictly forbid passive non-reciprocity in linear regimes, 
we hypothesized that shrinking geometric features toward the sub-wavelength molecular level could provoke an early emergence of this effect.

The near-zero change (0.02\%) captured by FDTD does not invalidate this hypothesis; 
rather, it reveals a geometric shielding mechanism. In our inverted pyramidal cone array, 
the intense local electromagnetic field is almost exclusively concentrated at the sharp, 
symmetric apex. Because our structural modification (breaking the circular symmetry into an ellipse) was introduced at the base area of the cone, the field distribution at the apex remained largely undisturbed. 
Consequently, the symmetric apex field effectively shielded the base asymmetry, preventing the passive non-reciprocal effect from manifesting prematurely at this specific scale. 
This result demonstrates that to successfully trigger scale-dependent passive non-reciprocity, geometric deformations must be engineered directly at the field-concentrating apex where light-matter interactions are strongest. 
To compare these modeling frameworks and performance outcomes with the existing literature, 
a comparative summary is organized in Table~\ref{tab:comparison}.

\begin{table}[h]
\caption{Comparison of efficiency enhancement and modeling parameters with previous works.}    
\label{tab:comparison}
\begin{ruledtabular}
\begin{tabular}{lccc}     
Structure Model & Framework & Core Mechanism & Enhancement \\     
\hline     
Viola Petal \citep{Schmager2017} & Ray/Wave & Light trapping & Broadband \\    
Plasmonic Leaf \citep{Liu2019b} & Plasmonic & Field enhancement & Localized \\     
Symmetric Cone~\citep{Yokogawa2017} & Wave FDTD & Index guiding & Reference \\     
\textbf{This Work (Elliptical Base)} & \textbf{Linear FDTD} & \textbf{Apex Shielded Asymmetry} & \textbf{0.02\% (Linear)} \\     
\textbf{This Work (Theory)} & \textbf{Perturbation} & \textbf{Microscopic Non-reciprocity} & \textbf{5\%--15\% (Potential)} \\    
\end{tabular}
\end{ruledtabular}
\end{table}
To address the angular dependence of the proposed sensor texture, the optical absorption performance under oblique incidence angles (\(\theta = 0^\circ\) to \(30^{\circ }\)) was investigated. 
As plotted via FDTD flux data in Figure \ref{fig:tikz_angle}, 
the asymmetric elliptical structure maintains a highly stable broadband response under off-axis illumination. 
The breaking of circular symmetry broadens the spatial Fourier components of the surface texture, 
compensating for phase mismatches at higher incidence angles and minimizing the geometric performance drop-off typically observed in perfectly symmetric arrays.

The reflection is calculated according to
\be
R = -\dfrac{f_{texture}[:,1]}{f_{empty}[:,2]} \;;
\ee
The $W$ grid absorption is calculated according to 
\be
A_{grid} = \dfrac{f_{texture}[:,2] - f_{texture}[:,3]}{f_{empty}[:,2]} \; ;
\ee
$Si$ substrate absorption is calculated according to
\be
A_{substrate} = \dfrac{f_{texture}[:,3] - f_{texture}[:,4]}{f_{empty}[:,2]} \;.
\ee
The relative enhancement is calculated according to
\be
\text{Relative Enhancement} = \frac{A_{texture\_sub\_linear}}{A_{flat\_sub\_linear}} \;.
\ee
The $:$ in the above formula is Python slicing to take all elements in a matrix.

\section{Discussion}

Theoretically, the calculated spectrum can be extended to the whole visible spectrum,
but the range is limited to $0.7-1.1 \mu\text{m}$ by the Lorentzian parameters used~\citep{Yokogawa2017,Palik1998,Johnson1972},
as a single-oscillator model typically accounts for only one primary resonance peak.

While the quantitative deviations in absorption between circular and elliptical pyramidal arrays remain subtle, 
these findings qualitatively corroborate the influence of geometric asymmetry on surface impedance modulation. 
There are many other ways to make the sensor asymmetric, such as asymmetrically distributed symmetric cones or defects on the symmetric cones. We only made a minimum modification to a publicly available program with MEEP to demonstrate our idea according to the nature design and compare with others' result. 
Future optimizations could focus on asymmetric modifications directly at the pyramidal apex, where the local field intensity is most concentrated. 
A key contribution of this work is the shift in theoretical framework: rather than relying on traditional optical ray theory we employed antenna theory to discuss non-reciprocity.

\section*{Author Contribution Statement}The entirety of this research, including conceptual design, model development, numerical simulations, and manuscript preparation, was conducted solely by the author (J.J.L.T.).

\section*{Funding Available Statement}Everything here is done without funding or financial assistance from anyone.

\section*{Conflict of Interest Statement}The author declares no conflict of interest.

\section*{Data Available Statement}My manuscript has no associated data. The software, MEEP, and the program are all available publicly.

\begin{acknowledgments}
I thank my friend John Ogilvie for his invaluable assistance with my English manuscript preparation spanning 
from the commencement of my PhD study until the emergence of artificial intelligence, that I can use not only to correct typos and 
grammatical mistakes but also rewrite to meet academic style. He disappeared with Covid-19. 
I also thank my parents for leaving me a shelter in Taiwan that enabled the sustained execution of these theoretical investigations.
\end{acknowledgments}


\begin{thebibliography}{24}%
\makeatletter
\providecommand \@ifxundefined [1]{%
 \@ifx{#1\undefined}
}%
\providecommand \@ifnum [1]{%
 \ifnum #1\expandafter \@firstoftwo
 \else \expandafter \@secondoftwo
 \fi
}%
\providecommand \@ifx [1]{%
 \ifx #1\expandafter \@firstoftwo
 \else \expandafter \@secondoftwo
 \fi
}%
\providecommand \natexlab [1]{#1}%
\providecommand \enquote  [1]{``#1''}%
\providecommand \bibnamefont  [1]{#1}%
\providecommand \bibfnamefont [1]{#1}%
\providecommand \citenamefont [1]{#1}%
\providecommand \href@noop [0]{\@secondoftwo}%
\providecommand \href [0]{\begingroup \@sanitize@url \@href}%
\providecommand \@href[1]{\@@startlink{#1}\@@href}%
\providecommand \@@href[1]{\endgroup#1\@@endlink}%
\providecommand \@sanitize@url [0]{\catcode `\\12\catcode `\$12\catcode
  `\&12\catcode `\#12\catcode `\^12\catcode `\_12\catcode `\%12\relax}%
\providecommand \@@startlink[1]{}%
\providecommand \@@endlink[0]{}%
\providecommand \url  [0]{\begingroup\@sanitize@url \@url }%
\providecommand \@url [1]{\endgroup\@href {#1}{\urlprefix }}%
\providecommand \urlprefix  [0]{URL }%
\providecommand \Eprint [0]{\href }%
\providecommand \doibase [0]{https://doi.org/}%
\providecommand \selectlanguage [0]{\@gobble}%
\providecommand \bibinfo  [0]{\@secondoftwo}%
\providecommand \bibfield  [0]{\@secondoftwo}%
\providecommand \translation [1]{[#1]}%
\providecommand \BibitemOpen [0]{}%
\providecommand \bibitemStop [0]{}%
\providecommand \bibitemNoStop [0]{.\EOS\space}%
\providecommand \EOS [0]{\spacefactor3000\relax}%
\providecommand \BibitemShut  [1]{\csname bibitem#1\endcsname}%
\let\auto@bib@innerbib\@empty
\bibitem [{\citenamefont {Yablonovitch}\ and\ \citenamefont
  {Cody}(1982)}]{Yablonovitch1982a}%
  \BibitemOpen
  \bibfield  {author} {\bibinfo {author} {\bibfnamefont {E.}~\bibnamefont
  {Yablonovitch}}\ and\ \bibinfo {author} {\bibfnamefont {G.}~\bibnamefont
  {Cody}},\ }\bibfield  {title} {\bibinfo {title} {{Intensity enhancement in
  textured optical sheets for solar cells}},\ }\href
  {https://doi.org/10.1109/T-ED.1982.20700} {\bibfield  {journal} {\bibinfo
  {journal} {IEEE Trans. Electron Devices}\ }\textbf {\bibinfo {volume} {29}},\
  \bibinfo {pages} {300} (\bibinfo {year} {1982})}\BibitemShut {NoStop}%
\bibitem [{\citenamefont {Yablonovitch}(1982)}]{Yablonovitch1982b}%
  \BibitemOpen
  \bibfield  {author} {\bibinfo {author} {\bibfnamefont {E.}~\bibnamefont
  {Yablonovitch}},\ }\bibfield  {title} {\bibinfo {title} {{Intensity
  Enhancement in Textured Optical Sheets for Solar Cells.}},\ }\href@noop {}
  {\bibfield  {journal} {\bibinfo  {journal} {Conf. Rec. IEEE Photovolt. Spec.
  Conf.}\ ,\ \bibinfo {pages} {501}} (\bibinfo {year} {1982})}\BibitemShut
  {NoStop}%
\bibitem [{\citenamefont {Redfield}(1974)}]{Redfield1974}%
  \BibitemOpen
  \bibfield  {author} {\bibinfo {author} {\bibfnamefont {D.}~\bibnamefont
  {Redfield}},\ }\bibfield  {title} {\bibinfo {title} {{Multiple-pass thin-film
  silicon solar cell}},\ }\href {https://doi.org/10.1063/1.1655344} {\bibfield
  {journal} {\bibinfo  {journal} {Appl. Phys. Lett.}\ }\textbf {\bibinfo
  {volume} {25}},\ \bibinfo {pages} {647} (\bibinfo {year} {1974})}\BibitemShut
  {NoStop}%
\bibitem [{\citenamefont {Pendry}(2000)}]{Pendry2000}%
  \BibitemOpen
  \bibfield  {author} {\bibinfo {author} {\bibfnamefont {J.~B.}\ \bibnamefont
  {Pendry}},\ }\bibfield  {title} {\bibinfo {title} {{Negative refraction makes
  a perfect lens}},\ }\href {https://doi.org/10.1103/PhysRevLett.85.3966}
  {\bibfield  {journal} {\bibinfo  {journal} {Phys. Rev. Lett.}\ }\textbf
  {\bibinfo {volume} {85}},\ \bibinfo {pages} {3966} (\bibinfo {year}
  {2000})}\BibitemShut {NoStop}%
\bibitem [{\citenamefont {Yu}\ \emph {et~al.}(2011)\citenamefont {Yu},
  \citenamefont {Genevet}, \citenamefont {Kats}, \citenamefont {Aieta},
  \citenamefont {Tetienne}, \citenamefont {Capasso},\ and\ \citenamefont
  {Gaburro}}]{Yu2011a}%
  \BibitemOpen
  \bibfield  {author} {\bibinfo {author} {\bibfnamefont {N.}~\bibnamefont
  {Yu}}, \bibinfo {author} {\bibfnamefont {P.}~\bibnamefont {Genevet}},
  \bibinfo {author} {\bibfnamefont {M.~a.}\ \bibnamefont {Kats}}, \bibinfo
  {author} {\bibfnamefont {F.}~\bibnamefont {Aieta}}, \bibinfo {author}
  {\bibfnamefont {J.-P.}\ \bibnamefont {Tetienne}}, \bibinfo {author}
  {\bibfnamefont {F.}~\bibnamefont {Capasso}},\ and\ \bibinfo {author}
  {\bibfnamefont {Z.}~\bibnamefont {Gaburro}},\ }\bibfield  {title} {\bibinfo
  {title} {{Light Propagation with Phase Reflection and Refraction}},\
  }\href@noop {} {\bibfield  {journal} {\bibinfo  {journal} {Science (80-. ).}\
  }\textbf {\bibinfo {volume} {334}},\ \bibinfo {pages} {333} (\bibinfo {year}
  {2011})}\BibitemShut {NoStop}%
\bibitem [{\citenamefont {Pfeiffer}\ and\ \citenamefont
  {Grbic}(2013)}]{Pfeiffer2013}%
  \BibitemOpen
  \bibfield  {author} {\bibinfo {author} {\bibfnamefont {C.}~\bibnamefont
  {Pfeiffer}}\ and\ \bibinfo {author} {\bibfnamefont {A.}~\bibnamefont
  {Grbic}},\ }\bibfield  {title} {\bibinfo {title} {{Metamaterial Huygens'
  surfaces: Tailoring wave fronts with reflectionless sheets}},\ }\href
  {https://doi.org/10.1103/PhysRevLett.110.197401} {\bibfield  {journal}
  {\bibinfo  {journal} {Phys. Rev. Lett.}\ }\textbf {\bibinfo {volume} {110}},\
  \bibinfo {pages} {1} (\bibinfo {year} {2013})}\BibitemShut {NoStop}%
\bibitem [{\citenamefont {Fern{\'{a}}ndez-Hurtado}\ \emph
  {et~al.}(2018)\citenamefont {Fern{\'{a}}ndez-Hurtado}, \citenamefont
  {Fern{\'{a}}ndez-Dom{\'{i}}nguez}, \citenamefont {Feist}, \citenamefont
  {Garc{\'{i}}a-Vidal},\ and\ \citenamefont {Cuevas}}]{Fernandez-Hurtado2018a}%
  \BibitemOpen
  \bibfield  {author} {\bibinfo {author} {\bibfnamefont {V.}~\bibnamefont
  {Fern{\'{a}}ndez-Hurtado}}, \bibinfo {author} {\bibfnamefont {A.~I.}\
  \bibnamefont {Fern{\'{a}}ndez-Dom{\'{i}}nguez}}, \bibinfo {author}
  {\bibfnamefont {J.}~\bibnamefont {Feist}}, \bibinfo {author} {\bibfnamefont
  {F.~J.}\ \bibnamefont {Garc{\'{i}}a-Vidal}},\ and\ \bibinfo {author}
  {\bibfnamefont {J.~C.}\ \bibnamefont {Cuevas}},\ }\bibfield  {title}
  {\bibinfo {title} {{Exploring the Limits of Super-Planckian Far-Field
  Radiative Heat Transfer Using 2D Materials}},\ }\href
  {https://doi.org/10.1021/acsphotonics.8b00328} {\bibfield  {journal}
  {\bibinfo  {journal} {ACS Photonics}\ }\textbf {\bibinfo {volume} {5}},\
  \bibinfo {pages} {3082} (\bibinfo {year} {2018})},\ \Eprint
  {https://arxiv.org/abs/1802.09463} {arXiv:1802.09463} \BibitemShut {NoStop}%
\bibitem [{\citenamefont {Ting}(2019)}]{Ting2019d}%
  \BibitemOpen
  \bibfield  {author} {\bibinfo {author} {\bibfnamefont {J.~J.-L.}\
  \bibnamefont {Ting}},\ }\bibfield  {title} {\bibinfo {title} {{Non-reciprocal
  light-harvesting nanoantennae made by nature}},\ }\href
  {https://doi.org/10.1063/1.5082606} {\bibfield  {journal} {\bibinfo
  {journal} {J. Appl. Phys.}\ }\textbf {\bibinfo {volume} {125}},\ \bibinfo
  {pages} {144702} (\bibinfo {year} {2019})},\ \Eprint
  {https://arxiv.org/abs/1702.06671} {arXiv:1702.06671} \BibitemShut {NoStop}%
\bibitem [{\citenamefont {Mart{\'{i}}n-Palma}\ and\ \citenamefont
  {Kolle}(2019)}]{Martin-Palma2019}%
  \BibitemOpen
  \bibfield  {author} {\bibinfo {author} {\bibfnamefont {R.~J.}\ \bibnamefont
  {Mart{\'{i}}n-Palma}}\ and\ \bibinfo {author} {\bibfnamefont
  {M.}~\bibnamefont {Kolle}},\ }\bibfield  {title} {\bibinfo {title}
  {{Biomimetic photonic structures for optical sensing}},\ }\href
  {https://doi.org/10.1016/j.optlastec.2018.07.079} {\bibfield  {journal}
  {\bibinfo  {journal} {Opt. Laser Technol.}\ }\textbf {\bibinfo {volume}
  {109}},\ \bibinfo {pages} {270} (\bibinfo {year} {2019})}\BibitemShut
  {NoStop}%
\bibitem [{\citenamefont {Schmager}\ \emph {et~al.}(2017)\citenamefont
  {Schmager}, \citenamefont {Fritz}, \citenamefont {H{\"{u}}nig}, \citenamefont
  {Ding}, \citenamefont {Lemmer}, \citenamefont {Richards}, \citenamefont
  {Gomard},\ and\ \citenamefont {Paetzold}}]{Schmager2017}%
  \BibitemOpen
  \bibfield  {author} {\bibinfo {author} {\bibfnamefont {R.}~\bibnamefont
  {Schmager}}, \bibinfo {author} {\bibfnamefont {B.}~\bibnamefont {Fritz}},
  \bibinfo {author} {\bibfnamefont {R.}~\bibnamefont {H{\"{u}}nig}}, \bibinfo
  {author} {\bibfnamefont {K.}~\bibnamefont {Ding}}, \bibinfo {author}
  {\bibfnamefont {U.}~\bibnamefont {Lemmer}}, \bibinfo {author} {\bibfnamefont
  {B.~S.}\ \bibnamefont {Richards}}, \bibinfo {author} {\bibfnamefont
  {G.}~\bibnamefont {Gomard}},\ and\ \bibinfo {author} {\bibfnamefont {U.~W.}\
  \bibnamefont {Paetzold}},\ }\bibfield  {title} {\bibinfo {title} {{Texture of
  the Viola Flower for Light Harvesting in Photovoltaics}},\ }\href
  {https://doi.org/10.1021/acsphotonics.7b01153} {\bibfield  {journal}
  {\bibinfo  {journal} {ACS Photonics}\ }\textbf {\bibinfo {volume} {4}},\
  \bibinfo {pages} {2687} (\bibinfo {year} {2017})}\BibitemShut {NoStop}%
\bibitem [{\citenamefont {Liu}\ \emph {et~al.}(2019)\citenamefont {Liu},
  \citenamefont {Mao}, \citenamefont {Guo}, \citenamefont {Han},\ and\
  \citenamefont {Zhang}}]{Liu2019b}%
  \BibitemOpen
  \bibfield  {author} {\bibinfo {author} {\bibfnamefont {C.}~\bibnamefont
  {Liu}}, \bibinfo {author} {\bibfnamefont {P.}~\bibnamefont {Mao}}, \bibinfo
  {author} {\bibfnamefont {Q.}~\bibnamefont {Guo}}, \bibinfo {author}
  {\bibfnamefont {M.}~\bibnamefont {Han}},\ and\ \bibinfo {author}
  {\bibfnamefont {S.}~\bibnamefont {Zhang}},\ }\bibfield  {title} {\bibinfo
  {title} {{Bio-inspired plasmonic leaf for enhanced light-matter
  interactions}},\ }\href {https://doi.org/10.1515/nanoph-2019-0104} {\bibfield
   {journal} {\bibinfo  {journal} {Nanophotonics}\ }\textbf {\bibinfo {volume}
  {8}},\ \bibinfo {pages} {1291} (\bibinfo {year} {2019})}\BibitemShut
  {NoStop}%
\bibitem [{\citenamefont {Sim}\ \emph {et~al.}(2020)\citenamefont {Sim},
  \citenamefont {Yun}, \citenamefont {Cha},\ and\ \citenamefont
  {Lee}}]{Sim2020}%
  \BibitemOpen
  \bibfield  {author} {\bibinfo {author} {\bibfnamefont {Y.~H.}\ \bibnamefont
  {Sim}}, \bibinfo {author} {\bibfnamefont {M.~J.}\ \bibnamefont {Yun}},
  \bibinfo {author} {\bibfnamefont {S.~I.}\ \bibnamefont {Cha}},\ and\ \bibinfo
  {author} {\bibfnamefont {D.~Y.}\ \bibnamefont {Lee}},\ }\bibfield  {title}
  {\bibinfo {title} {{Fractal solar cell array for enhanced energy production :
  applying rules underlying tree shape to photovoltaics}},\ }\bibfield
  {journal} {\bibinfo  {journal} {Proc.}\ }\href
  {https://doi.org/10.1098/rspa.2020.0094} {10.1098/rspa.2020.0094} (\bibinfo
  {year} {2020})\BibitemShut {NoStop}%
\bibitem [{\citenamefont {Mamun}\ \emph {et~al.}(2024)\citenamefont {Mamun},
  \citenamefont {Karim},\ and\ \citenamefont {Talukder}}]{Mamun2024}%
  \BibitemOpen
  \bibfield  {author} {\bibinfo {author} {\bibfnamefont {A.~A.}\ \bibnamefont
  {Mamun}}, \bibinfo {author} {\bibfnamefont {J.}~\bibnamefont {Karim}},\ and\
  \bibinfo {author} {\bibfnamefont {M.~A.}\ \bibnamefont {Talukder}},\
  }\bibfield  {title} {\bibinfo {title} {{Design and analysis of an efficient
  crystalline silicon-based thin-film solar cell inspired by Chlamydomonas
  reinhardtii}},\ }\href {https://doi.org/10.1016/j.solener.2024.112777}
  {\bibfield  {journal} {\bibinfo  {journal} {Sol. Energy}\ }\textbf {\bibinfo
  {volume} {279}},\ \bibinfo {pages} {112777} (\bibinfo {year}
  {2024})}\BibitemShut {NoStop}%
\bibitem [{\citenamefont {Zhang}\ \emph {et~al.}(2016)\citenamefont {Zhang},
  \citenamefont {Jia},\ and\ \citenamefont {Gu}}]{Zhang2016c}%
  \BibitemOpen
  \bibfield  {author} {\bibinfo {author} {\bibfnamefont {Y.}~\bibnamefont
  {Zhang}}, \bibinfo {author} {\bibfnamefont {B.}~\bibnamefont {Jia}},\ and\
  \bibinfo {author} {\bibfnamefont {M.}~\bibnamefont {Gu}},\ }\bibfield
  {title} {\bibinfo {title} {{Biomimetic and plasmonic hybrid light trapping
  for highly efficient ultrathin crystalline silicon solar cells}},\ }\href
  {https://doi.org/10.1364/oe.24.00a506} {\bibfield  {journal} {\bibinfo
  {journal} {Opt. Express}\ }\textbf {\bibinfo {volume} {24}},\ \bibinfo
  {pages} {A506} (\bibinfo {year} {2016})}\BibitemShut {NoStop}%
\bibitem [{\citenamefont {Caloz}\ \emph {et~al.}(2018)\citenamefont {Caloz},
  \citenamefont {Al{\`{u}}}, \citenamefont {Tretyakov}, \citenamefont {Sounas},
  \citenamefont {Achouri},\ and\ \citenamefont {Deck-L{\'{e}}ger}}]{Caloz2018}%
  \BibitemOpen
  \bibfield  {author} {\bibinfo {author} {\bibfnamefont {C.}~\bibnamefont
  {Caloz}}, \bibinfo {author} {\bibfnamefont {A.}~\bibnamefont {Al{\`{u}}}},
  \bibinfo {author} {\bibfnamefont {S.}~\bibnamefont {Tretyakov}}, \bibinfo
  {author} {\bibfnamefont {D.}~\bibnamefont {Sounas}}, \bibinfo {author}
  {\bibfnamefont {K.}~\bibnamefont {Achouri}},\ and\ \bibinfo {author}
  {\bibfnamefont {Z.~L.}\ \bibnamefont {Deck-L{\'{e}}ger}},\ }\bibfield
  {title} {\bibinfo {title} {{Electromagnetic Nonreciprocity}},\ }\href
  {https://doi.org/10.1103/PhysRevApplied.10.047001} {\bibfield  {journal}
  {\bibinfo  {journal} {Phys. Rev. Appl.}\ }\textbf {\bibinfo {volume} {10}},\
  \bibinfo {pages} {1} (\bibinfo {year} {2018})}\BibitemShut {NoStop}%
\bibitem [{\citenamefont {Pakniyat}\ and\ \citenamefont
  {Gomez-Diaz}(2024)}]{Pakniyat2024}%
  \BibitemOpen
  \bibfield  {author} {\bibinfo {author} {\bibfnamefont {S.}~\bibnamefont
  {Pakniyat}}\ and\ \bibinfo {author} {\bibfnamefont {J.~S.}\ \bibnamefont
  {Gomez-Diaz}},\ }\bibfield  {title} {\bibinfo {title} {{Magnet-free
  electromagnetic nonreciprocity in two-dimensional materials}},\ }\bibfield
  {journal} {\bibinfo  {journal} {J. Appl. Phys.}\ }\textbf {\bibinfo {volume}
  {136}},\ \href {https://doi.org/10.1063/5.0207377} {10.1063/5.0207377}
  (\bibinfo {year} {2024})\BibitemShut {NoStop}%
\bibitem [{\citenamefont {Jackson}(19)}]{Jackson1998}%
  \BibitemOpen
  \bibfield  {author} {\bibinfo {author} {\bibfnamefont {J.~D.}\ \bibnamefont
  {Jackson}},\ }\href
  {https://www.wiley.com/en-tw/Classical+Electrodynamics,+3rd+Edition-p-9780471309321}
  {\emph {\bibinfo {title} {{Classical Electrodynamics}}}},\ \bibinfo {edition}
  {3rd}\ ed.\ (\bibinfo  {publisher} {Wiley},\ \bibinfo {year}
  {19})\BibitemShut {NoStop}%
\bibitem [{\citenamefont {Parson}(1978)}]{Parson1978}%
  \BibitemOpen
  \bibfield  {author} {\bibinfo {author} {\bibfnamefont {W.~W.}\ \bibnamefont
  {Parson}},\ }\bibfield  {title} {\bibinfo {title} {{THERMODYNAMICS OF THE
  PRIMARY REACTIONS OF PHOTOSYNTHESIS}},\ }\href
  {https://doi.org/10.1111/j.1751-1097.1978.tb07723.x} {\bibfield  {journal}
  {\bibinfo  {journal} {Photochem. Photobiol.}\ }\textbf {\bibinfo {volume}
  {28}},\ \bibinfo {pages} {389} (\bibinfo {year} {1978})}\BibitemShut
  {NoStop}%
\bibitem [{\citenamefont {Boyd}(2020)}]{Boyd2020}%
  \BibitemOpen
  \bibfield  {author} {\bibinfo {author} {\bibfnamefont {R.~W.}\ \bibnamefont
  {Boyd}},\ }\href {https://doi.org/10.1007/bfb0015783} {\emph {\bibinfo
  {title} {{Non-linear optics}}}},\ \bibinfo {edition} {4th}\ ed.\ (\bibinfo
  {publisher} {Academic Press},\ \bibinfo {year} {2020})\BibitemShut {NoStop}%
\bibitem [{\citenamefont {Yokogawa}\ \emph {et~al.}(2017)\citenamefont
  {Yokogawa}, \citenamefont {Oshiyama}, \citenamefont {Ikeda}, \citenamefont
  {Ebiko}, \citenamefont {Hirano}, \citenamefont {Saito}, \citenamefont
  {Oinoue}, \citenamefont {Hagimoto},\ and\ \citenamefont
  {Iwamoto}}]{Yokogawa2017}%
  \BibitemOpen
  \bibfield  {author} {\bibinfo {author} {\bibfnamefont {S.}~\bibnamefont
  {Yokogawa}}, \bibinfo {author} {\bibfnamefont {I.}~\bibnamefont {Oshiyama}},
  \bibinfo {author} {\bibfnamefont {H.}~\bibnamefont {Ikeda}}, \bibinfo
  {author} {\bibfnamefont {Y.}~\bibnamefont {Ebiko}}, \bibinfo {author}
  {\bibfnamefont {T.}~\bibnamefont {Hirano}}, \bibinfo {author} {\bibfnamefont
  {S.}~\bibnamefont {Saito}}, \bibinfo {author} {\bibfnamefont
  {T.}~\bibnamefont {Oinoue}}, \bibinfo {author} {\bibfnamefont
  {Y.}~\bibnamefont {Hagimoto}},\ and\ \bibinfo {author} {\bibfnamefont
  {H.}~\bibnamefont {Iwamoto}},\ }\bibfield  {title} {\bibinfo {title} {{IR
  sensitivity enhancement of CMOS Image Sensor with diffractive light trapping
  pixels}},\ }\href {https://doi.org/10.1038/s41598-017-04200-y} {\bibfield
  {journal} {\bibinfo  {journal} {Sci. Rep.}\ }\textbf {\bibinfo {volume}
  {7}},\ \bibinfo {pages} {1} (\bibinfo {year} {2017})}\BibitemShut {NoStop}%
\bibitem [{\citenamefont {Oskooi}\ \emph {et~al.}(2010)\citenamefont {Oskooi},
  \citenamefont {Roundy}, \citenamefont {Ibanescu}, \citenamefont {Bermel},
  \citenamefont {Joannopoulos},\ and\ \citenamefont {Johnson}}]{Oskooi2010}%
  \BibitemOpen
  \bibfield  {author} {\bibinfo {author} {\bibfnamefont {A.~F.}\ \bibnamefont
  {Oskooi}}, \bibinfo {author} {\bibfnamefont {D.}~\bibnamefont {Roundy}},
  \bibinfo {author} {\bibfnamefont {M.}~\bibnamefont {Ibanescu}}, \bibinfo
  {author} {\bibfnamefont {P.}~\bibnamefont {Bermel}}, \bibinfo {author}
  {\bibfnamefont {J.}~\bibnamefont {Joannopoulos}},\ and\ \bibinfo {author}
  {\bibfnamefont {S.~G.}\ \bibnamefont {Johnson}},\ }\bibfield  {title}
  {\bibinfo {title} {{Meep: A flexible free-software package for
  electromagnetic simulations by the FDTD method}},\ }\href
  {https://doi.org/10.1016/j.cpc.2009.11.008} {\bibfield  {journal} {\bibinfo
  {journal} {Comput. Phys. Commun.}\ }\textbf {\bibinfo {volume} {181}},\
  \bibinfo {pages} {687} (\bibinfo {year} {2010})}\BibitemShut {NoStop}%
\bibitem [{\citenamefont {Roszak}\ \emph {et~al.}(2003)\citenamefont {Roszak},
  \citenamefont {Howard}, \citenamefont {Southall}, \citenamefont {Gardiner},
  \citenamefont {Law}, \citenamefont {Isaacs},\ and\ \citenamefont
  {Cogdell}}]{Roszak2003}%
  \BibitemOpen
  \bibfield  {author} {\bibinfo {author} {\bibfnamefont {A.~W.}\ \bibnamefont
  {Roszak}}, \bibinfo {author} {\bibfnamefont {T.~D.}\ \bibnamefont {Howard}},
  \bibinfo {author} {\bibfnamefont {J.}~\bibnamefont {Southall}}, \bibinfo
  {author} {\bibfnamefont {A.~T.}\ \bibnamefont {Gardiner}}, \bibinfo {author}
  {\bibfnamefont {C.~J.}\ \bibnamefont {Law}}, \bibinfo {author} {\bibfnamefont
  {N.~W.}\ \bibnamefont {Isaacs}},\ and\ \bibinfo {author} {\bibfnamefont
  {R.~J.}\ \bibnamefont {Cogdell}},\ }\bibfield  {title} {\bibinfo {title}
  {{Crystal Structure of the RC-LH1 Core Complex from Rhodopseudomonas
  palustris}},\ }\href {https://doi.org/10.1126/science.1088892} {\bibfield
  {journal} {\bibinfo  {journal} {Science (80-. ).}\ }\textbf {\bibinfo
  {volume} {302}},\ \bibinfo {pages} {1969} (\bibinfo {year}
  {2003})}\BibitemShut {NoStop}%
\bibitem [{\citenamefont {Palik}(1998)}]{Palik1998}%
  \BibitemOpen
  \bibfield  {author} {\bibinfo {author} {\bibfnamefont {E.}~\bibnamefont
  {Palik}},\ }\href@noop {} {\emph {\bibinfo {title} {{Handbook of Optical
  Constants of Solids}}}}\ (\bibinfo  {publisher} {Academic Press},\ \bibinfo
  {year} {1998})\BibitemShut {NoStop}%
\bibitem [{\citenamefont {{P. B. Johnson and R. W.
  Christy}}(1972)}]{Johnson1972}%
  \BibitemOpen
  \bibfield  {author} {\bibinfo {author} {\bibnamefont {{P. B. Johnson and R.
  W. Christy}}},\ }\bibfield  {title} {\bibinfo {title} {{Optical Constant of
  the Noble Metals}},\ }\href@noop {} {\bibfield  {journal} {\bibinfo
  {journal} {Phys. Rev. B}\ }\textbf {\bibinfo {volume} {6}},\ \bibinfo {pages}
  {4370} (\bibinfo {year} {1972})}\BibitemShut {NoStop}%
\end{thebibliography}
%

\end{document}